\documentclass[reprint,showpacs,amsmath,amssymb,aps,prb]{revtex4-1}
\usepackage{graphicx}
\usepackage{dcolumn}
\usepackage{bm}
\usepackage{hyperref}
\usepackage[mathlines]{lineno}

\begin{document}

\title{Quantum Anomalous Hall Effect of Magnetic Topological Insulator Thin Films by Phase Boundary Engineering}
\author{Kiminori Hattori}
\affiliation{Graduate School of Engineering Science, Osaka University, Toyonaka, Osaka 560-8531, Japan}

\begin{abstract}
We generalize the topological phase diagram of magnetic topological insulator (TI) thin films in an extended parameter space comprising out-of-plane (OP) and in-plane (IP) exchange fields in the presence of structural inversion asymmetry (SIA) while taking a generic orbital-dependent spin coupling allowed for TIs into consideration.
The results show that an IP field substantially deforms phase boundaries and generically induces the quantum anomalous Hall (QAH) effect.
For symmetric spin coupling, extremely weak OP and IP exchange fields create the QAH state by tuning SIA with a gate bias.
For antisymmetric coupling, the QAH phase is absent without a strong enough IP field.
These findings demonstrate that in the thin-film regime, engineering the phase boundary is a key process to efficiently realize and manipulate the QAH effect for nondissipative electronic applications.
\end{abstract}

\maketitle

\section{Introduction}

A thin-film topological insulator (TI) provides a planar platform that supports a variety of topological states and is suitable for exploiting related topological effects in actual devices.
In a thin-film configuration, surfaces states of a three-dimensional (3D) TI~\cite{ref:1, ref:2, ref:3, ref:4} lying on the opposite surfaces hybridize to create a tunneling gap.
Then, the system alternates between the topologically-trivial normal insulator (NI) phase and the nontrivial quantum spin Hall (QSH) phase, depending on the film thickness.\cite{ref:5, ref:6, ref:7, ref:8, ref:9, ref:10, ref:11, ref:12}
The QSH state consists of two copies of quantum Hall (QH) states with opposite Chern numbers, arranged so that time-reversal symmetry is preserved.\cite{ref:13, ref:14, ref:15, ref:16}
In a finite system with open boundaries, there exists on each edge a Kramers pair of chiral edge states inside the gap.
The helical edge mode is essentially robust against time-reversal-invariant nonmagnetic perturbations.
However, magnetic disorder due to random magnetic impurities is detrimental to the QSH transport since they give rise to intraedge backscattering.\cite{ref:16}

This vulnerability can be eliminated by incorporating ferromagnetism into the system by magnetic doping.\cite{ref:17, ref:18, ref:19, ref:20, ref:21}
A strong enough exchange field normal to the surface drives the NI or QSH phase into the quantum anomalous Hall (QAH) phase, where the QH effect is activated in either of two subsystems and suppressed in the other.\cite{ref:22, ref:23, ref:24, ref:25, ref:26}
The QAH phase hosts a single chiral edge mode on each edge, for which backscattering is completely forbidden.
Hence, the QAH insulator exhibits perfect conduction that is insensitive to magnetic and nonmagnetic disorder as long as the gap remains intact.\cite{ref:16}
The QAH effect is experimentally observed in the Bi$_2$Se$_3$ family of materials~\cite{ref:27, ref:28, ref:29, ref:30}, and is theoretically proposed to occur in other related systems~\cite{ref:31, ref:32, ref:33, ref:34, ref:35}.

There are extensive studies in this context, where an out-of-plane (OP) and orbital-independent exchange field constitutes a main control parameter to induce the QAH effect for a given hybridization gap.
Here, we generalize the topological phase diagram of magnetic TI thin films by considering an in-plane (IP) exchange field in addition to the OP field, structural inversion asymmetry (SIA) due to an OP electric field, and generic orbital-dependent exchange coupling for the Dirac model describing 3DTIs.
It is shown in the extended parameter space that an IP exchange field generically induces the QAH effect by appreciably deforming phase boundaries, leading to an unconventional scheme to control the topological state of a thin-film TI.

\section{Theory}

The $4 \times 4$ Dirac Hamiltonian describing nonmagnetic 3DTIs is given in momentum space by
\begin{equation}
\label{eq:1}
{\mathcal{H}_0}=D({\mathbf{k}})+{\mathbf{A}}({\mathbf{k}}) \cdot {\tau_x}{\bm{\sigma}}+B({\mathbf{k}}){\tau_z} ,
\end{equation}
where $D={D_1}{k_z}^2+{D_2}{k^2}$, ${\mathbf{A}}=({A_2}{k_x},{A_2}{k_y},{A_1}{k_z})$, $B={B_0}-{B_1}{k_z}^2-{B_2}{k^2}$, and ${k^2}={k_x}^2+{k_y}^2$.\cite{ref:1, ref:2}
The parameters in these definitions are estimated for the Bi$_2$Se$_3$ family.\cite{ref:1, ref:2}
${\sigma_\mu}$ and ${\tau_\mu}$ with $\mu=x,y,z$ represent the Pauli matrices in spin and orbital spaces, respectively.
To formulate the exchange interaction with magnetic moments in a homogeneous phase, we consider the generic form of Zeeman-type coupling allowed in the Dirac Hamiltonian~\cite{ref:2, ref:20}
\begin{equation}
\label{eq:2}
{\mathcal{H}_m}=({\mathbf{m}}+{\mathbf{n}}{\tau_z}) \cdot {\bm{\sigma}} .
\end{equation}
The exchange field (in energy units) is composed of the symmetric part ${\mathbf{m}}$ and the antisymmetric part ${\mathbf{n}}$, by which the possible orbital dependence of spin coupling is fully describable.
Thus, the total Hamiltonian consists of $\mathcal{H}={\mathcal{H}_0}+{\mathcal{H}_m}$.
See Appendix A for more details.
We first analyze the contribution due to the symmetric field ${\mathbf{m}}$, which accords with the conventional approach used in the study of magnetic TIs.
The effect of the antisymmetric counterpart ${\mathbf{n}}$ will be addressed later.

Diagonalizing ${\mathcal{H}_0}$ under an open boundary condition and projecting $\mathcal{H}$ onto the subspace of surface states,\cite{ref:1, ref:2, ref:7, ref:8, ref:11, ref:12, ref:23, ref:24, ref:25} we obtain the effective low-energy Hamiltonian in a thin-film configuration.
The result is summarized as $H={H_0}+{H_m}$ with
\begin{gather}
\label{eq:3}
{H_0}={\varepsilon}({\mathbf{k}})+{\gamma}{s_z}{({\bm{\sigma}} \times {\mathbf{k}})_z}+t({\mathbf{k}}){s_x}+{m_z}{\sigma_z} ,\\[3pt]
\label{eq:4}
{H_m}=(I+J{s_x}){{\mathbf{m}}_{||}} \cdot {\bm{\sigma}} ,
\end{gather}
in the basis constituted of top and bottom surface states $\{ t_{\uparrow},t_{\downarrow},b_{\uparrow },b_{\downarrow } \}$.
Here, ${s_\mu}$ denotes the Pauli matrix in orbital space spanned by $\{ t,b \}$.
${H_m}$ describes the exchange interaction associated with the IP field ${{\mathbf{m}}_{||}}=({m_x},{m_y},0)$.
The contribution due to the OP field ${m_z}$ is incorporated into ${H_0}$ for convenience.
${H_0}$ preserves rotational symmetry while it is broken by ${H_m}$.
The parameters are given by $\varepsilon={E_0}+D{k^2}$, $t={\Delta}-B{k^2}$, $D={D_2}-I{B_2}$, $B=J{B_2}$, $I=\frac{\left\langle t \right|{\tau_z}\left| t \right\rangle+\left\langle b \right|{\tau_z}\left| b \right\rangle}{2}$, and $J=\operatorname{Re}\left\langle t \right|{\tau_z}\left| b \right\rangle$.
By definition, $\left|J\right|$ reflects intersurface mixing, and is essentially augmented with a reduction in the film thickness ${L_z}$.
For instance, $(I,J)=(0.062,-0.7)$ at ${L_z}=2.3{\text{nm}}$ and $(0.12,-0.35)$ at 3nm for Bi$_2$Se$_3$,\cite{ref:12} whereas $I={D_1}/{B_1}=0.13$ and $J=0$ in the thick-film limit ${L_z}\to\infty$.\cite{ref:1, ref:2, ref:8, ref:25}
This suggests that in the thin-film regime, the $J$ term is dominant in ${H_m}$.

The orthogonal transformation $\left|{\varphi_{1,2}}\right\rangle=(\left| t \right\rangle \pm \left| b \right\rangle)/\sqrt 2$ arranges ${H_0}$ into a block diagonal form.
The eigenenergies of two blocks labeled by $p = \pm 1$ are written as $E_r^{(p)}=\varepsilon+r{E^{(p)}}$ and ${E^{(p)}}=\sqrt{{\gamma^2}{k^2}+{{(t+p{m_z})}^2}}$, where $r = \pm 1$ refer to the conduction and valence bands, respectively.
The band gap ${E_g}=2\min\left|{\Delta \pm m_z}\right|$ closes at the $\Gamma$ point ($k=0$) when ${m_z}^2={\Delta^2}$.
On the other hand, it is shown from the Chern numbers of the associated four bands that block $p$ becomes nontrivial when $(\Delta+p{m_z})B>0$.\cite{ref:7, ref:8, ref:11, ref:12, ref:23, ref:24}
As a result, the system enters into the QAH phase for ${m_z}^2>{\Delta^2}$.
If this criterion is not fulfilled, the system lies in the QSH phase for $\Delta B>0$ or the NI phase for $\Delta B<0$.
These three phases are topologically distinct in terms of the Chern numbers and cannot smoothly connect without closing the gap.

The total Hamiltonian $H={H_0}+{H_m}$ can be also analytically diagonalized for $I=0$.
Defining the orthogonal two wavevectors ${{\mathbf{k}}_\ell}={{\mathbf{m}}_{||}}({{\mathbf{m}}_{||}} \cdot {\mathbf{k}})/{m_{||}}^2$ and ${{\mathbf{k}}_t}={\mathbf{k}}-{{\mathbf{k}}_\ell}$, the gap function is expressed as
\begin{equation}
\label{eq:5}
{E^{(p)}}=\sqrt{{{(T+pM)}^2}+{\gamma^2}({k^2}-\eta{k_\ell}^2)} ,
\end{equation}
where $T=\sqrt{{t^2}+\eta{\gamma^2}{k_\ell}^2}$, $M=\sqrt{{m_z}^2+{J^2}{m_{||}}^2}$, and $\eta=\frac{J^2{m_{||}}^2}{{m_z}^2+J^2{m_{||}}^2}$.
It is found from Eq.~(\ref{eq:5}) that the gap closes when $T-M=(1-\eta){k_\ell}={k_t}=0$.
These coupled conditions reduce to $k=0$ and ${M^2}={\Delta^2}$ for ${m_z} \ne 0$, and predict that the QAH phase emerges in the region where ${M^2}>{\Delta^2}$.
The system otherwise belongs to the QSH phase for $\Delta B>0$ or the NI phase for $\Delta B <0$.
Thus, adding ${H_m}$ to ${H_0}$ is formally equivalent to replacing ${m_z}^2$ by ${M^2}$ in terms of gap closing, demonstrating that ${m_z}$ and ${m_{||}}$ equally drive the phase transition into the QAH state.
If ${m_z}=0$, the gap remains closed as long as ${J^2}{m_{||}}^2>{\Delta^2}$, and then the system becomes metallic.
From these considerations, we expect the Chern number to be formulated as
\begin{equation}
\label{eq:6}
C=-\operatorname{sgn}({m_z})\theta({m_z}^2+{J^2}{m_{||}}^2-{\Delta^2}) .
\end{equation}
It is worth noting that two QAH phases classified by $C= \pm 1$ touch for ${J^2}{m_{||}}^2>{\Delta^2}$ at ${m_z}= 0$ where the gap vanishes.
In the QSH phase, the bulk gap is open even for ${m_{||}} \ne 0$, although helical edge states formed in a finite system are gapped by intraedge mixing.
In this sense, the system under a small IP field is no longer in the ideal QSH state.
In the following, we nevertheless refer to it as a QSH system to avoid unnecessary and probably confusing nomenclature.
For more details, see Appendix B.

A similar phase transition arising from an IP field has been investigated previously for paramagnetic HgMnTe quantum wells,\cite{ref:36} although its physical mechanism remains unresolved.
To understand the underlying physics, we here analyze the effective Hamiltonian for a weak IP field.
The relevant Hamiltonian is formulated as $\tilde{H}={H_0}+{H_m}{(E_r^{(p)}-{H_0})^{-1}}{H_m}$ for an unperturbed eigenstate of energy $E_r^{(p)}$ following the second-order perturbation formalism.
$\tilde{H}$ is block diagonal in the basis chosen for ${H_0}$.
An identity term in each block is negligible since it has no effect on the band topology.
The leading terms in two blocks construct the rotationally-invariant effective Hamiltonian, expressed as
\begin{equation}
\label{eq:7}
\tilde{H}=\gamma{s_z}{(\bm{\sigma} \times {\mathbf{k}})_z}+\tilde{t}({\mathbf{k}}){s_x}+{\tilde{m}_z}{\sigma_z} ,
\end{equation}
yielding the renormalized parameters to second order in $k$; $\tilde{t}=\tilde{\Delta}-\tilde{B}{k^2}$, $\tilde{\Delta}=\Delta-{J^2}{m_{||}}^2/4\Delta$, $\tilde{B}=B(1+{J^2}{m_{||}}^2/4{\Delta^2})$, and $\tilde{m}_z={m_z}+{J^2}{m_{||}}^2/4{m_z}$.
It is easily shown from these formulae that the original QAH criterion ${m_z}^2>{\Delta^2}$ substantially changes into ${m_z}^2+{J^2}{m_{||}}^2>{\Delta^2}$ to second order in ${m_{||}}$.
The renormalization thus reproduces the conclusion drawn from gap closing.

It is also instructive to consider a generalized 3D parameter space $(\Delta,{m_z},{m_{||}})$.
Scanning a 2D section $(\Delta,{m_z})$ along ${m_{||}}$, the renormalization due to ${m_{||}}$ is viewed as a continuous deformation of the QAH phase boundary, which consists of an equilateral hyperbola and a line segment joining two vertices at $(\pm J{m_{||}},0)$.
As ${m_{||}}$ increases, the QAH effect is activated in the otherwise inactive region by deforming the phase boundary outward, i.e., ${m_{||}}$ intrinsically extends the QAH phase.
However, the renormalization is no longer valid in the thick-film limit where $\Delta=J=0$.
In this limit, top and bottom surface states are completely decoupled, and ${\mathbf{m}}_{||}$ merely shifts each Dirac dispersion in the opposite direction by an amount ${s_z}I({\mathbf{m}}_{||} \times \hat{\mathbf{z}})/\gamma$.
This is a clear distinction showing that the renormalization and the consequent phase-boundary deformation are characteristic of thin-film TIs.

\section{Numerical Calculation}

To examine quantitative aspects, we have performed numerical calculation of the Chern number using the Kubo formula $C=\frac{1}{2\pi}\sum_{n \in \text{occ}}\int_{\text{BZ}}d\mathbf{k}\Omega_n$ for a periodic system,\cite{ref:37} where the summation runs over all occupied states at half filling, and the integration is over the Brillouin zone.
The momentum-space Berry curvature is given by
\begin{equation}
\label{eq:8}
{\Omega_n}(\mathbf{k})=2\operatorname{Im}\sum\limits_{m \ne n}{\frac{{\left\langle n \right|\tfrac{{\partial H}}{{\partial {k_x}}}\left| m \right\rangle \left\langle m \right|\tfrac{{\partial H}}{{\partial {k_y}}}\left| n \right\rangle }}{{{{({E_n}-{E_m})}^2}}}} ,
\end{equation}
where $H\left|n\right\rangle={E_n}\left|n\right\rangle$.
For simplicity, we set ${\mathbf{m}}_{||}=({m_x},0,0)$ without loss of generality because ${H_0}$ is rotationally invariant.
The film thicknesses were chosen as ${L_z}=2.3{\text{nm}}$ and 3nm, for which equally $\left|\Delta\right|=20.2{\text{meV}}$.
In the absence of exchange coupling, the 2.3nm- and 3nm-thick systems are in the NI phase ($\Delta B<0$) and in the QSH phase ($\Delta B>0$), respectively.
Further details are described in our previous paper.\cite{ref:12}
In the calculation, the numerical values of $I$ and $J$ evaluated from surface wavefunctions were adopted.

\begin{figure}
\centering
\includegraphics{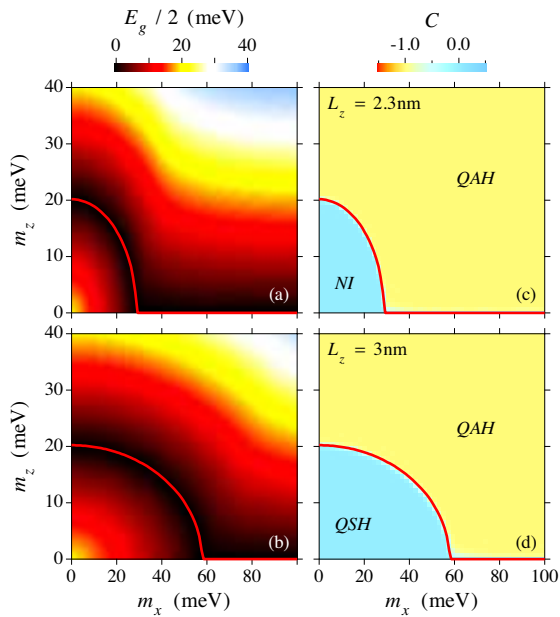}
\caption{(Color online) [(a) and (b)] Band gap ${E_g}$ and [(c) and (d)] Chern number $C$ numerically computed as a function of symmetric exchange fields ${m_x}$ and ${m_z}$.
In the calculation, SIA is neglected ($V=0$).
Solid lines represent the phase boundaries by which the QAH phase is separated from the NI phase for [(a) and (c)] ${L_z}=2.3{\text{nm}}$ and from the QSH phase for [(b) and (d)] ${L_z}=3{\text{nm}}$.}
\label{fig:1}
\end{figure}

The numerical results are shown in Fig.~\ref{fig:1}.
As expected, the gap-closing line along ${m_z}^2+{J^2}{m_{||}}^2={\Delta^2}$ for ${J^2}{m_{||}}^2<{\Delta^2}$ and ${m_z}=0$ for ${J^2}{m_{||}}^2>{\Delta^2}$ separates the QAH phase featured by $C=-\operatorname{sgn}({m_z})$ from the NI or QSH phases carrying $C=0$.
In the region where ${m_z}^2<{\Delta^2}$, a strong enough ${m_{||}}$ brings about the topological phase transition into the QAH phase.
The formation of the QAH state is independently verified from chiral edge states in an open-boundary system and the resulting quantization of Hall resistance (not shown).
The system stays in the QAH phase for ${m_z}^2>{\Delta^2}$ irrespective of ${m_x}$.
In this regime, the band gap continuously increases with increasing ${m_x}$.
The QAH state is topologically protected and essentially immune to external perturbations unless the gap collapses.
In this sense, the QAH effect is stabilized by adding an IP field through the gap augmentation.

Next, we will take SIA stemming from an OP electric field into consideration.
The relevant Hamiltonian is simply given by~\cite{ref:8, ref:11, ref:24}
\begin{equation}
\label{eq:9}
{H_v}=V{s_z} .
\end{equation}
In the absence of ${H_m}$, the total Hamiltonian $H={H_0}+{H_v}$ is easily diagonalized.
The gap function is given by ${E^{(p)}}=\sqrt{{{(T+pM)}^2}+(1-\eta){\gamma^2}{k^2}}$, where $T=\sqrt{{t^2}+{V^2}}$, $M=\sqrt{{m_z}^2+\eta{\gamma^2}{k^2}}$, and $\eta=\frac{V^2}{t^2+V^2}$.
It is found from the gap-closing condition $T-M=(1-\eta)k=0$ that the QAH phase emerges when ${m_z}^2>{\Delta^2}+V^2$, indicating that SIA tends to suppress the QAH effect.\cite{ref:24}
This can be explicitly shown in $(\Delta,{m_z})$ space from the phase-boundary deformation due to $V$.
The QSH phase exists for $\Delta B>0$ within the interval ${V^2}-{V_0}^2<{m_z}^2<{\Delta^2}+{V^2}$.
Here, ${V_0}=\left|\gamma\right|{k_0}$, and ${k_0}=\sqrt{\Delta/B}$ is the critical momentum at which $\eta=1$.\cite{ref:11, ref:24}
The NI phase appears otherwise, i.e., ${m_z}^2<{V^2}-{V_0}^2$ for $\Delta B>0$ and ${m_z}^2<{\Delta^2}+{V^2}$ for $\Delta B<0$.

\begin{figure}
\centering
\includegraphics{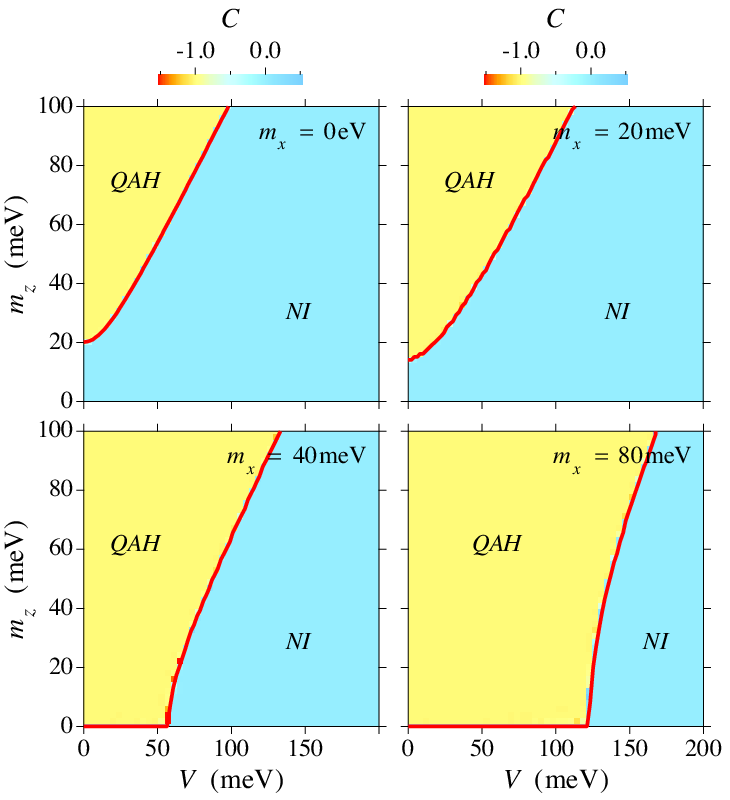}
\caption{(Color online) Numerical values of Chern number for ${L_z}=2.3{\text{nm}}$ as a function of $V$ and ${m_z}$ at (a) ${m_x}=0$, (b) 20meV, (c) 40meV, and (d) 80meV.
Solid lines represent the phase boundaries where the gap closes.}
\label{fig:2}
\end{figure}

\begin{figure}
\centering
\includegraphics{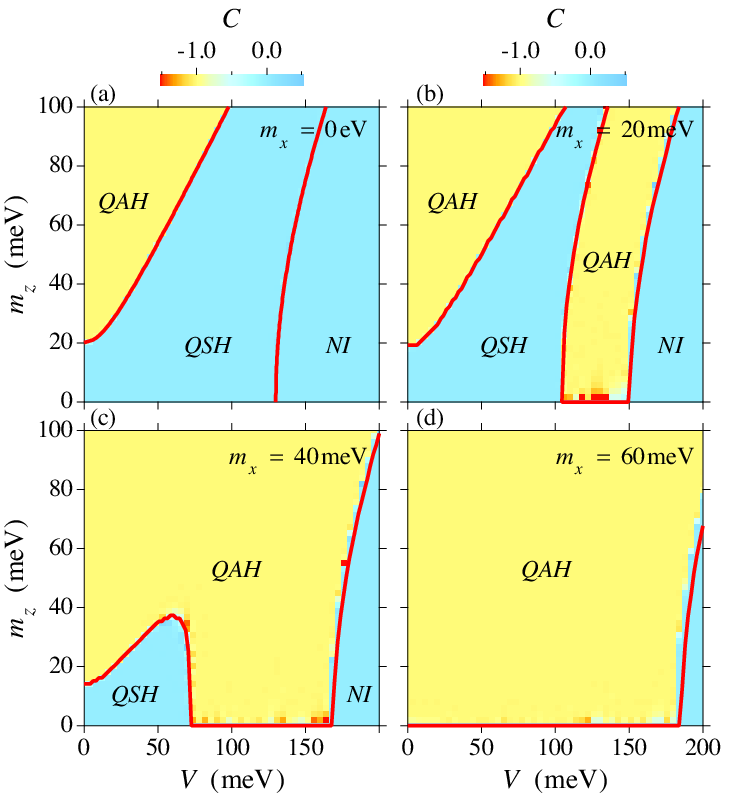}
\caption{(Color online) Numerical values of Chern number for ${L_z}=3{\text{nm}}$ as a function of $V$ and ${m_z}$ at (a) ${m_x}=0$, (b) 20meV, (c) 40meV, and (d) 60meV.
Solid lines represent the phase boundaries where the gap closes.}
\label{fig:3}
\end{figure}

Figure~\ref{fig:2} summarizes the numerical results for ${L_z}=2.3{\text{nm}}$ in the $(V,{m_z})$ plane for various ${m_x}$ values.
Solid lines in the figure indicate the points where the gap vanishes.
As is clearly seen, the QAH phase gradually extends with increasing ${m_x}$, and eventually covers the entire parameter space.
A similar behavior is also observed for ${L_z}=3{\text{nm}}$ in Fig.~\ref{fig:3}.
It is confirmed from these observations that even in the presence of SIA, an IP field largely deforms phase boundaries and induces the QAH effect independently of the initial topology.

\begin{figure}
\centering
\includegraphics{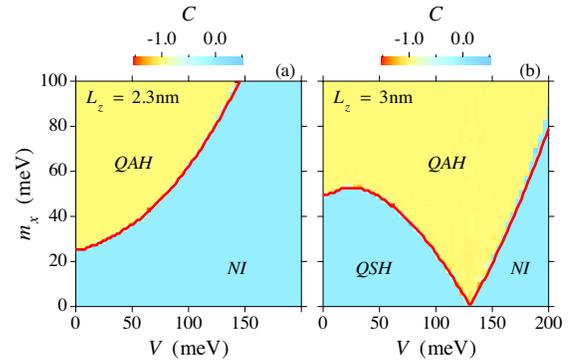}
\caption{(Color online) Numerical values of Chern number at ${m_z}=10{\text{meV}}$ as a function of $V$ and ${m_x}$ for (a) ${L_z}=2.3{\text{nm}}$ and (b) ${L_z}=3{\text{nm}}$.
Solid lines represent the phase boundaries where the gap closes.}
\label{fig:4}
\end{figure}

A remarkable feature captured in Fig.~\ref{fig:3} is penetration of the QAH phase through the original QSH-NI boundary in the presence of a small ${m_x}$.
To elucidate this peculiar behavior more quantitatively, the phase diagram in $(V,{m_x})$ space at ${m_z}=10{\text{meV}}$ ($<\left|\Delta\right|$) is displayed in Fig.~\ref{fig:4}.
In this diagram, the phase boundary ${m_x}(V)$ amounts to the critical field to generate the QAH effect.
For ${L_z}=2.3{\text{nm}}$, ${m_x}(V)$ monotonically increases as a function of $V$ from the finite lower bound given by ${m_x}(0)=\sqrt{(\Delta^2-{m_z}^2)/J^2}$.
On the other hand, ${m_x}(V)$ decreases along the QSH-QAH boundary and subsequently increases along the NI-QAH boundary for ${L_z}=3{\text{nm}}$.
Importantly, ${m_x}(V)$ vanishes at the intermediate critical point where these three phases contact.
In terms of the analytical results described above, such a triple point appears for $\Delta B>0$ at ${m_{||}}=0$ and ${V^2}={V_0}^2+{m_z}^2$, indicating that the QAH effect is inducible by extremely small ${m_z}$ and ${m_{||}}$ in a thin-film geometry.
This property presents a striking contrast to the prerequisite ${m_z}^2>{\Delta^2}+{V^2}$ for ${m_{||}}=0$,\cite{ref:24} which implies that a larger gap $\left|\Delta\right|$ due to a stronger intersurface mixing obstructs the phase transition into the QAH state.
The difficulty incidental to thickness reduction is circumvented by introducing a weak IP field.
It is also noticeable that around the triple point, the topological state of a thin-film TI is very sensitive to $V$ and is therefore tunable by a moderate gate bias.
The tunablity is intrinsic to the thin-film regime and is not easy to realize in thick films.

\section{Antisymmetric Exchange Coupling}

Finally, we examine the effects of antisymmetric exchange interaction.
Spin coupling of this type adds the term
\begin{equation}
\label{eq:10}
{H_n}=(I+J{s_x}){n_z}{\sigma_z}+{{\mathbf{n}}_{||}} \cdot {\bm{\sigma}} ,
\end{equation}
to the thin-film Hamiltonian.
The antisymmetric fields $I{n_z}$ and ${\mathbf{n}}_{||}$ can be absorbed into definitions of the symmetric counterparts ${m_z}$ and $I{\mathbf{m}}_{||}$, respectively.
The remaining term related to $J{n_z}$ is a new contribution.
Similar to ${m_z}+I{n_z}$, $J{n_z}$ breaks time-reversal and reflection symmetries.
The symmetry breaking is necessary for the QAH effect.\cite{ref:38}
Nevertheless, $J{n_z}$ primarily impedes the QAH effect.
This can be explicitly shown by assuming ${\mathbf{m}}_{||}={\mathbf{n}}_{||}=0$.
In this simplified situation, $J{n_z}$ gives rise to only an energy splitting between two blocks, ${E_0}^{(p)}={E_0}+pJ{n_z}$, and consequently the system becomes metallic when $J{n_z}$ is sufficiently large.\cite{ref:20}
The system remains in an insulating state for $\left|J{n_z}\right|<\max{(\left|\Delta\right|,\left|{{m_z}+I{n_z}}\right|)}$, whereas band inversion (gap closing and reopening) occurs in either of the two blocks when $\left|\Delta\right|<\left|{m_z+In_z}\right|$.
Combining these two conditions, we reach the effective QAH criterion $\left|Jn_z\right|<\left|{m_z+In_z}\right|$, which conflicts with the relation $\left|J\right|>I$ intrinsic to thin films if $m_z=0$.
In thick films, $I>\left|J\right| \approx 0$ so that the surface QAH effect is realizable even for ${m_z}=0$.
However, edge states are no longer robust to external perturbations because of mixing with nonchiral side-surface modes.\cite{ref:29}

\begin{figure}
\centering
\includegraphics{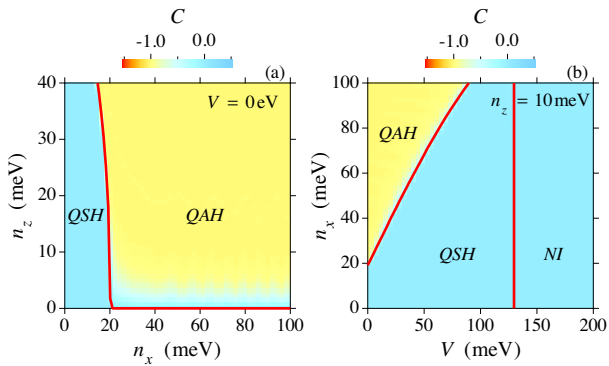}
\caption{(Color online) Numerical values of Chern number $C$ for ${L_z}=3{\text{nm}}$ in the case of antisymmetric exchange coupling: (a) $C({n_x},{n_z})$ at $V=0$ and (b) $C(V,{n_x})$ at ${n_z}=10{\text{meV}}$.
Solid lines represent the phase boundaries where the gap closes.}
\label{fig:5}
\end{figure}

The disappearance of the QAH effect in the thin-film regime is reversed by introducing an IP field ${n_{||}}$, as shown in Fig.~\ref{fig:5}.
Fundamentally, this effect is accountable for by the block mixing due to ${n_{||}}$, which substantially changes the QAH criterion into ${J^2}{n_z}^2+{n_{||}}^2>{\Delta^2}$ for $\mathbf{m}=0$.
As demonstrated above, SIA tends to quench the QAH effect for symmetric coupling.
This is also valid for antisymmetric coupling.
In this case, it is shown from the analytical diagonalization for $V \ne 0$ and ${n_{||}}=0$ that the QAH phase is absent; the system lies in either the QSH phase for ${V_0}^2>{V^2}+{J^2}{n_z}^2$ and $\Delta B>0$ or the NI phase otherwise.
However, a sufficiently large ${n_{||}}$ regenerates the QAH effect in the presence of SIA, although the triple point is not located around ${n_{||}}=0$ for a small ${n_z}$.
Thus, a nonzero IP field is essential to realize the QAH effect, particularly for antisymmetric coupling.
This observation is important in realistic situations since $\left|{n_z}\right|$ is estimated to be comparable to, or larger than, $\left|{m_z}\right|$ in a certain class of magnetic TIs.\cite{ref:20}

\section{Summary}

In summary, we have generalized the topological phase diagram of magnetic TI thin films by considering IP and OP exchange fields, SIA induced by an OP electric field, and symmetric and antisymmetric spin couplings with magnetization, on the basis of the Chern number computed in a multidimensional parameter space.
The results demonstrate that the phase-boundary deformation due to an IP exchange field generically induces the QAH effect.
For symmetric coupling, the QAH state is created by minimal OP and IP exchange fields in the proximity of the QSH-QAH-NI triple point.
For antisymmetric coupling, the QAH phase is absent without a strong enough IP field.

Even when magnetocrystalline anisotropy is perpendicular to the film, as observed for a large class of magnetic TI materials,\cite{ref:17, ref:19, ref:20, ref:21} an IP exchange field can be generated by applying an external magnetic field ${B_{||}}$ parallel to the film, which rotates the magnetization by an angle $\theta={\sin^{-1}}{B_{||}}/{B_K}$, where ${B_K}$ denotes the anisotropy field.
The auxiliary external field is suitable to separate exchange or Zeeman coupling from orbital effects in experiments.
Moreover, in terms of the anisotropy energy to second order $E={K_1}\sin^2\theta+{K_2}\sin^4\theta$, the magnetization canted by an angle $\theta=\sin^{-1}\sqrt{-{K_1}/2{K_2}}$ to the easy axis may form a ground state and stabilize with no external field by virtue of shape or surface anisotropy in a thin-film geometry.\cite{ref:39, ref:40}

\appendix

\section{Bulk Hamiltonian}

Here, we compare in more detail the bulk Hamiltonians, ${\mathcal{H}_0}$ and ${\mathcal{H}_m}$, to those presented in Refs. 1 and 2. ${\mathcal{H}_0}$ is identical to Eq. (1) in Ref. 1, as well as to Eq. (19) in Ref. 2 to second order in $k$.
They are equally represented as
\begin{equation*}
{\mathcal{H}_0}=D({\mathbf{k}})+{A_1}{\Gamma_3}{k_z}+{A_2}({\Gamma_1}{k_x}+{\Gamma_2}{k_y})+B({\mathbf{k}}){\Gamma_5} ,
\end{equation*}
where the $\Gamma$ matrices are defined as ${\Gamma_1}={\tau_x}{\sigma_x}$, ${\Gamma_2}={\tau_x}{\sigma_y}$, ${\Gamma_3}={\tau_x}{\sigma_z}$, ${\Gamma_4}={\tau_y}$ and ${\Gamma_5}={\tau_z}$.
${\mathcal{H}_0}$ is unitary equivalent to Eq. (16) in Ref. 2, expressed as
\begin{equation*}
{\mathcal{H'}_0}=D({\mathbf{k}})+{A_1}{\Gamma_4}{k_z}+{A_2}({\Gamma_1}{k_y}-{\Gamma_2}{k_x})+B({\mathbf{k}}){\Gamma_5} ,
\end{equation*}
as explicitly shown by ${\mathcal{H'}_0}={U_1}^\dag {\mathcal{H}_0}{U_1}$ with ${U_1}$ defined by Eq. (18) in Ref. 2.

On the other hand, applying this transformation to ${\mathcal{H}_m}=({\mathbf{m}}+{\mathbf{n}}{\tau_z}) \cdot {\bm{\sigma}}$ yields
\begin{equation*}
{\mathcal{H'}_m}=({m_z}+{n_z}{\tau_z}){\sigma_z}+({{\mathbf{n}}_{||}}+{{\mathbf{m}}_{||}}{\tau_z}) \cdot {\bm{\sigma}} .
\end{equation*}	
Thus, a symmetric (antisymmetric) IP field in this paper corresponds to an antisymmetric (symmetric) one for the basis chosen in Ref. 2, while symmetric and antisymmetric OP fields are not switched by this transformation.

\section{QSH phase}

In the absence of ${m_{||}}$, the QSH state in thin-film TIs is simply defined by the opposite Chern numbers of two blocks classified by $p=\pm 1$, which correspond to the eigenvalues of $P={s_x}{\sigma_z}$.
Since $[P,{H_0}]=0$, $p$ is a conserved quantity.\cite{ref:12}
However, it should be noted that the QSH state is not protected by the $P$-symmetry $P{H_0}{P^{-1}}={H_0}$. This is obvious from the phase transition into the QAH state due to a strong enough OP field.

As clearly shown in Fig. 1 (b), the bulk gap remains open in the QSH phase even for $m_{||} \ne 0$.
The observation implies the nontrivial bulk topology persists, since the Chern numbers cannot change for two insulating states adiabatically connected without closing the gap.
Analytically, this can be shown from the QSH criteria in the presence of a small $m_{||}$.
Following the effective Hamiltonian, Eq. (7), the system belongs to the QSH phase as long as ${\tilde m_z}^2 < {\tilde \Delta}^2$ and ${\tilde \Delta} {\tilde B}>0$.
These criteria are entirely verified in the QSH phase shown in Fig. 1 (d).
Based on Eqs. (3), (4) and (8), the opposite Chern numbers ${C_1} = -{C_2}$ are numerically confirmed for two valence bands (${E_1} \le {E_2}$) in the QSH phase, as shown in Fig. 6.
\begin{figure}
\centering
\includegraphics{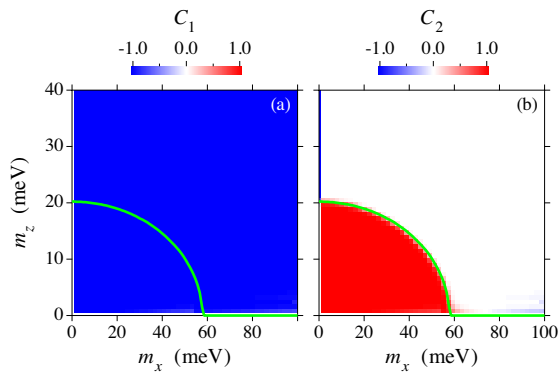}
\caption{(Color online) Numerical values of Chern numbers (a) ${C_1}$ and (b) ${C_2}$ for two valence bands (${E_1} \le {E_2}$).
Solid lines represent the phase boundaries where the gap closes.
Parameters are identical to those of Fig. 1 (d), where $C = {C_1} + {C_2}$ is shown.}
\label{fig:6}
\end{figure}
Note that gap closing and reopening change the Chern number of the associated higher band.

However, the resulting QSH effect is rather complicated in a finite system with open boundaries.
Figure 7 shows the energy spectra computed in a strip geometry of width $L_y$ with a periodic boundary condition in the $x$ direction and an open boundary condition in the $y$ direction.
\begin{figure}
\centering
\includegraphics{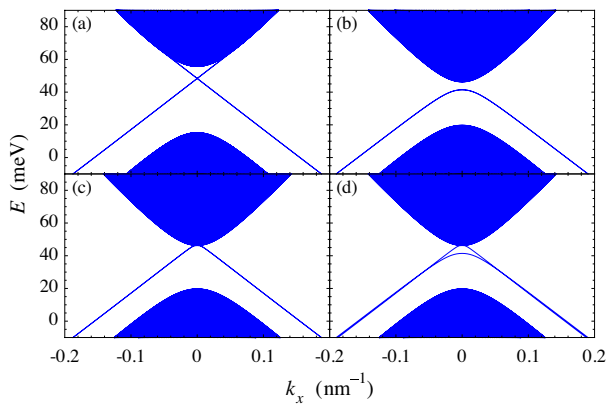}
\caption{(Color online) Energy dispersions computed in a strip geometry for (a) ${m_x} = 0$, (b) ${m_x} = m = 20{\text{meV}}$, (c) ${m_x}(y) = m \theta (\frac{{{L_y}}}{2} - w - \left| y \right|)$ and (d) ${m_x}(y) = m \theta (\frac{{{L_y}}}{2} - w - y)$.
Parameters are ${L_z} = 3{\text{nm}}$, ${L_y} = 2000{\text{nm}}$, ${m_y} = {m_z} = 0$ and $w = 30{\text{nm}}$.}
\label{fig:7}
\end{figure}
In the calculation, parameters are chosen to be ${L_z}=3{\text{nm}}$, ${L_y}=2000{\text{nm}}$ and ${m_y} = {m_z} = 0$.
In the finite system, gapless helical edge modes exist for ${m_x}=0$ [Fig. 7 (a)], while the edge modes are gapped by ${m_x} \equiv m = 20{\text{meV}}$ [Fig. 7 (b)].

To examine the gap-opening mechanism, we have performed the additional calculations that assume a nonuniform field ${m_x}(y)$ partially vanishing in narrow regions adjacent to the boundaries at $y= \pm {L_y}/2$.
The width of the field-free region is set at $w = 30{\text{nm}}$, which is slightly larger than the exponential decay length of edge-state wavefunction $\ell = 26{\text{nm}}$ for ${m_x} = 0$.\cite{ref:12}
Note that interedge interaction is negligible since $\ell << {L_y}$.
Since $w << {L_y}$, the following two results are naturally expected.
(i) The edge gap is sustained even for ${m_x}(y)$ if it reflects the bulk topology.
(ii) The edge gap is suppressed for ${m_x}(y)$ if it arises from local intraedge interaction.
The numerical results support the latter.
The edge gap almost disappears for ${m_x}(y) = m \theta (\frac{{{L_y}}}{2} - w - \left| y \right|)$ [Fig. 7 (c)].
Moreover, for ${m_x}(y) = m \theta (\frac{{{L_y}}}{2} - w - y)$, the lower edge mode subjected to ${m_x} \ne 0$ is gapped while the upper edge mode little affected by ${m_x}$ tends to be gapless [Fig, 7 (d)].
These results confirm that in the QSH phase, intraedge mixing due to ${m_{||}}$ creates the edge gap in a finite system, whereas the nontrivial bulk topology survives.

\bibliography{ref}

\end{document}